\documentclass[twocolumn, 11pt]{article}
\usepackage{array, float}
\usepackage[left=18mm, right=18mm, top=26mm, bottom=24mm]{geometry}
\usepackage{newtxtext}
\usepackage{amsmath}
\usepackage{mathrsfs}
\usepackage{tabularx}
\usepackage{graphicx}
\graphicspath{{./fig/}}
\usepackage{dcolumn}
\usepackage{setspace}
\usepackage{pgfpages}
\makeatletter
\usepackage{dcolumn}
\usepackage{url}
\usepackage{comment}
\usepackage{lineno}
\usepackage{cite}

\usepackage{abstract}

\usepackage{setspace}
\usepackage[font=small, format=plain, labelfont=bf, up, textfont=normal, up, justification=justified,singlelinecheck=false]{caption}

\usepackage[affil-it]{authblk} 

\let\OLDthebibliography\thebibliography
\renewcommand\thebibliography[1]{
\OLDthebibliography{#1}
\setlength{\parskip}{0pt}
\setlength{\itemsep}{0pt plus 0.3ex}
}

\title{Symmetry breaking in the prisoner's dilemma on two-layer dynamic multiplex networks}
\author{Hirofumi Takesue\thanks{Electronic address: \texttt{hir.takesue@gmail.com}}}
\affil{Faculty of Political Science and Economics, Waseda University}
\date{}

\begin{document}

\twocolumn[

\maketitle

\begin{onecolabstract}
Understanding the role of network structure in the evolution of cooperation is a key research goal at the intersection between physics and biology. Recent studies have particularly focused on multiplex networks given that multiple social domains are interrelated and cannot be represented by single-layer networks. However, the role of network multiplexity is not fully understood when combined with another important network characteristic: network dynamics. In the present study, we investigated evolutionary prisoner's dilemma games played on dynamic two-layer multiplex networks in which the payoff combined across the two layers determined strategy evolution. In addition, we introduced network dynamics where agents can sever links with defecting neighbors and construct new links. Our simulation showed that link updating enhances cooperation but the resultant states are far from those of full cooperation. This modest enhancement in cooperation was related to symmetry breaking whereby the cooperation frequency in one layer disproportionately increased while that in the other layer remained the same or even diminished. However, this broken symmetry disappeared with sufficiently fast link updating. Our results show that the introduction of network dynamics enhances cooperation in the prisoner's dilemma as previously reported, but this enhancement is accompanied by significant asymmetry once network multiplexity is considered.
\\\\
\end{onecolabstract}
]
\saythanks

\section*{Introduction}
The origin of cooperation is an intriguing research topic at the intersection between physical and biological sciences \cite{Nowak2006, Szabo2007a, Roca2009, Perc2017, Tanimoto2019}; it has been analyzed using a mathematical framework known as evolutionary game theory. The core dilemma in the study of cooperation is the discrepancy between myopic rationality and social efficiency. Specifically, non-cooperators can avoid the costs of cooperation while enjoying the benefits of others' cooperation; thus, non-cooperation is advantageous to individual interests. Consequently, this free-riding leads to the prevalence of non-cooperation and lower social efficiency. 

Network reciprocity, where network structure supports the maintenance of cooperation, offers one potential solution to this dilemma. On the one hand, cooperators cannot survive in well-mixed populations because defectors can achieve larger payoffs on average by avoiding the cost of cooperation. On the other hand, a limited number of neighbors interacting in a network facilitate the formation of cooperative clusters, which enable cooperators to achieve higher fitness from the benefits of mutual cooperation. Since the seminal work of Nowak and May \cite{Nowak1992}, researchers have examined the effects of various network characteristics, including degree heterogeneity \cite{Santos2005, Santos2006b}, average degree \cite{Ohtsuki2006a, Grafen2007}, and assortativity \cite{Rong2007}, on cooperation. One study has also clarified the relationship between network reciprocity and a fundamental concept in evolutionary biology, namely inclusive fitness \cite{Grafen2007}. 

In the present study, we focus on the \textit{multiplexity} and \textit{dynamics} of networks in relation to cooperation. Multilayer networks that are not limited to multiplex networks are a key research focus in network science \cite{Boccaletti2014}. Understanding these types of network is crucial because multiple types of (social) activity are interrelated and should therefore be represented by networks with multiple layers. A seminal study showed that failure in one layer (e.g., in power networks) can lead to severe fragmentation in multiple layers (e.g., in Internet networks as well as power networks) \cite{Buldyrev2010}. In addition, multiplex networks show novel epidemic spreading patterns \cite{Granell2013} and contribute to robust diversity in culture formation models \cite{Battiston2017a}.

Multilayer networks are also vital to studying the evolution of cooperation \cite{Wang2015e}. One widely examined interdependency among network layers is payoff coupling whereby individuals' performances depend on the game payoff from multiple layers \cite{Wang2012d, Wang2013, Meng2015, Wang2016j, Allen2017a, Kleineberg2018} and the existence of an optimal interdependency level is indicated \cite{Wang2012e, Wang2013a}. More complex situations, in which the layers differed in the games conducted \cite{Santos2014, Wang2014f, Deng2018c, Xia2018} or topological characters \cite{Li2019e}, have also been studied. In addition, network layers can be coupled by other factors such as information about strategy frequency \cite{Szolnoki2013e}, imitation probability \cite{Liu2019j}, reputation \cite{Wang2017b}, social pressure \cite{Pereda2016}, or the selection of imitation partners \cite{Wang2014b}. Furthermore, several studies have demonstrated the coevolution of cooperation and interdependency among network layers \cite{Wang2014e, Wang2014g, Liu2018h, Yang2018f, Jia2019}. 

Multilayer networks have been found to support cooperation through incoherent behavior  whereby individuals adopt different strategies in different network layers; some studies suggest that cooperation enhancement can be attributed to this incoherent behavior rather than the consistent adoption of cooperation in multiple layers of the network \cite{Gomez-Gardenes2012, Matamalas2015}. Moreover, such disparity over networks appears not only at the individual level but also at the macroscopic level; interdependent networks show \textit{symmetry breaking} by which the overall cooperation levels in each layer diverge. Thus, although the same rules may be applied across the network, the extent of cooperation enhancement can differ across layers \cite{Jin2014, Battiston2017, Liu2019k}.

Network dynamics is another realistic network characteristic \cite{Gross2008} that researchers have investigated in relation to cooperation \cite{Perc2010}. In evolutionary games, network dynamics imply that the existence or duration of links between individuals depends on the individuals' attributes including their strategy. Both theoretical \cite{Zimmermann2004, Eguiluz2005, Pacheco2006a, Santos2006, Fu2007, Tanimoto2007, Fu2008, VanSegbroeck2008, Fu2009, Szolnoki2009b, Tanimoto2009, Wu2009c, Zhang2011, Li2013a, Yang2013a, Cong2014, Li2014a, Xu2014, Chen2016, Pinheiro2016, Wang2016f, Li2017b, Takesue2018} and experimental \cite{Fehl2011, Rand2011, Gallo2015} studies have demonstrated that network dynamics strongly enhance cooperation; however, overly fast link dynamics have been shown to hinder cooperation \cite{Fu2009, Shirado2013}. Researchers have also indicated that network dynamics are important to other types of cooperation-related phenomena such as fairness \cite{Gao2011, Takesue2017a, Takesue2020}. Recently, the role of network dynamics was elucidated further: sufficiently fast link updating in a network was shown to result in full cooperation in two-layer multiplex networks \cite{Yang2019d}.

Although independent lines of research on multiplex networks and network dynamics have provided valuable insights into the role of networks in the evolution of cooperation, their coupled effects have yet to be fully examined (a notable exception is Ref. \cite{Yang2019d}). To remedy this situation, here we examined the prisoner's dilemma game played on dynamic multiplex networks. Specifically, agents were located on networks with two layers (duplex networks) and played the prisoner's dilemma game with their neighbors. The two layers were mutually related because the payoff that accumulated over both layers determined the evolution of strategies. Links in each layer could be modified depending on the strategy adopted by agents. 

The results of our simulation showed that introducing link updating to the network increased cooperation but that the end result was far from full cooperation. The modest enhancement in cooperation is related to symmetry breaking whereby cooperation frequencies in one layer increase while those in another layer decrease or remain constant. However, we found that this broken symmetry disappeared once the speed of link updating became overly fast. In summary, our model shows that network dynamics support cooperation but are accompanied by nontrivial asymmetry once network multiplexity is considered.

\section*{Simulation model}
We consider a duplex network in which each $N$ agent occupies one node in both layers (please refer to Table \ref{tab_params} for the overview of parameters in the model). Each agent will participate in the prisoner's dilemma game with her direct neighbors (see Ref. \cite{Tanimoto2007a} for the overview of dilemma situations). In each layer, agents adopt one of two strategies: cooperation ($C$) or defection ($D$). Agent $i$'s strategy in layer $l$ is denoted by $s_i^{(l)}$. Through the interaction with agent $j$ on layer $l$, agent $i$ acquires payoff $\pi_{s_i^{(l)} s_j^{(l)}}$. In the prisoner's dilemma, the order of the four payoff values is $\pi_{DC} > \pi_{CC} > \pi_{DD} > \pi_{CD}$. Because defection results in a larger payoff regardless of a partner's decision, less profitable mutual defection tends to be realized. Following convention, we basically set the value of $\pi_{CC}$ ($\pi_{DD}$) to 1 (0) and controlled the harshness of social dilemmas through the values of $\pi_{DC}$ and $\pi_{CD}$. Because we assume multiplex networks, all agents occupy one node in each layer, and their overall payoff is determined by game interactions in two layers as explained below. 
\renewcommand{\arraystretch}{1.15}
\begin{table*}[tbp]
\centering
\caption{\small Overview of the parameters}
\begin{tabular}{ccc}
\hline
\hline
Symbol & Explanation & Range \\[0.1ex]
\hline
$N$ & Number of agents & 1000 (but see Fig. \ref{fig_rho_lm}) \\ 
$p$ & \begin{tabular}{c}Link generation probability of \\ Erd\H{o}s-R\'{e}nyi networks \end{tabular} & 0.01 (but see Fig. \ref{fig_rho_lm}) \\
$w$ & Probability of link updating & [0, 0.95] \\
$\mu$ & Mutation probability & 0.0001 \\
$\pi_{CC}$ & Payoff & 1 (but see Fig. \ref{fig_scale}) \\
$\pi_{CD}$ & Payoff & \begin{tabular}{c} [-0.975, -0.025] \\ (but see Fig. \ref{fig_scale}) \end{tabular} \\
$\pi_{DC}$ & Payoff & \begin{tabular}{c} [1.025, 1.975] \\ (but see Fig. \ref{fig_scale}) \end{tabular} \\
$\pi_{DD}$ & Payoff & 0 (but see Fig. \ref{fig_scale}) \\
\hline
\hline
\end{tabular}
\label{tab_params}
\end{table*}
\renewcommand{\arraystretch}{1}

In their initial states, agents are located on symmetrical Erd\H{o}s-R\'{e}nyi random networks in which the two layers share the same set of agents and edges. In other words, the links between each pair of nodes on one layer are generated with a probability of $p$ as in the same manner with one layer random networks; the other layer is a duplicate of this generated layer. One study \cite{Battiston2017} showed that link overlap between network layers enhances cooperation. Because symmetrical initial states maximize link overlap (and enhance cooperation), this initial state assures the conservative estimate of the positive impact of link updating on cooperation. Agents' strategies in each layer are randomly assigned; strategies in two layers may not be consistent. 

\begin{figure*}[tbp]
\centering
\vspace{5mm}
\includegraphics[width = 135mm, trim= 0 10 0 0]{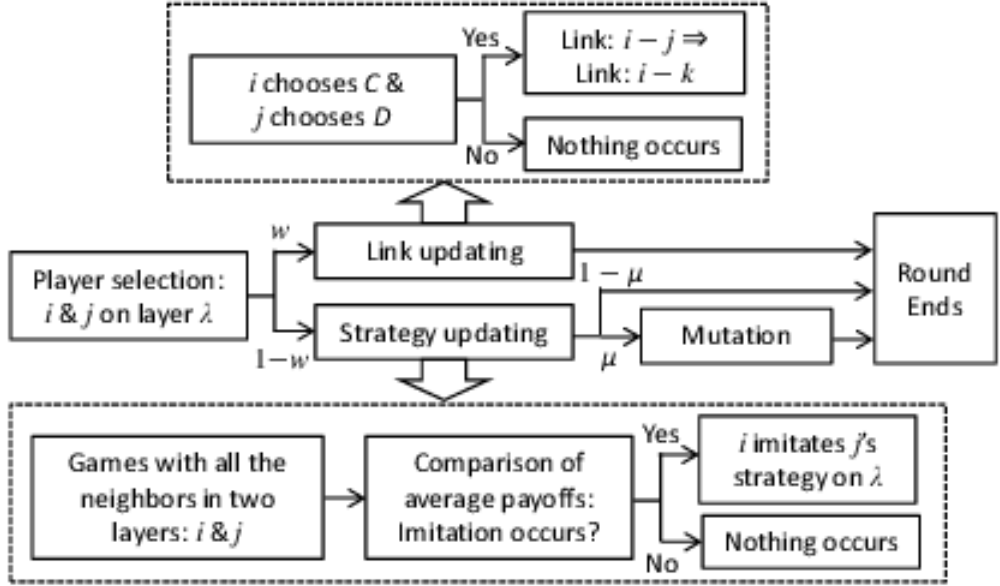}
\caption{\small Schematic overview of the simulation process.}
\label{fig_process}
\end{figure*}

Two types of events modify this system: strategy updating and link updating. In each round, one of these two events occurs; link updating occurs with a probability of $w$ whereas strategy updating occurs with a probability of $1-w$. This parameter controls the speed of the network dynamics relative to strategy evolution (please refer to Figure \ref{fig_process} for the schematic presentation of updating processes in one elementary time step). 

During strategy updating, agents may imitate their neighbor's strategy in a specific layer. In such an event, one layer ($\lambda$) and one \textit{link} in that layer ($e_{i, j}^{(\lambda)}$) are randomly selected; one of the connected agents is randomly selected to become a focal agent,  while the other becomes a role agent \cite{Fu2009, Takesue2019}. In the explanation provided here, agent $i$ becomes a focal agent while agent $j$ becomes a role agent. A study showed that this link-based rule suppresses cooperation than when determining a focal agent by the random selection of one \textit{node} from a whole population \cite{Takesue2019}, and we will examine how this rule works with multiplex networks. 

In this study, we adopt simple network interdependency and assume that agents' payoffs are determined by the interactions with all of their neighbors on both layers \cite{Gomez-Gardenes2012}. Formally, the focal agent's payoff is determined as follows: 
\begin{equation}
 \Pi_i = \sum_{l \in \left\{1,2\right\}}\sum_{k \in \mathcal{N}_i^{(l)}} \pi_{s_i^{(l)} s_k^{(l)}} / (z_i^{(1)} + z_i^{(2)}),
\end{equation}
where $\mathcal{N}_i^{(l)}$ is the set of the focal agent's neighbors in layer $l$. The accumulated payoff is regularized by the sum of the agent's degree in two layers ($z_i^{(1)}$ + $z_i^{(2)}$). The role agent earns her payoff, $\Pi_j$, in the same manner. Adopting average payoffs means that achieving large degree ($z_i^{(1)}$ + $z_i^{(2)}$) does not necessarily lead to higher fitness while the difference in degree between two layers can influence the behavior of agents through the different number of games and their payoff. This point will be discussed in reporting the simulation results. 

The focal agent may imitate the role agent's strategy on layer $\lambda$ when the role agent acquired larger payoff; the imitation probability is given by Wu et al. \cite{Wu2007} as follows:
\begin{equation}
 {\rm P}(s_i^{(\lambda)} \leftarrow s_j^{(\lambda)}) = \max \left\{(\Pi_j - \Pi_i) / (\pi_{DC} - \pi_{CD}), 0 \right\}.
\end{equation}
Here, payoff accumulation introduces interdependency between two layers but strategy transmission occurs only on the selected layer.

Strategy mutation occurs with a small probability ($\mu$) in strategy updating. In mutations, the focal agent ignores the results of the payoff-based imitation described above and instead adopts one strategy randomly. Mutation causes small perturbations in the system and prevents spurious frozen states. These small perturbations are known to have significant impacts on various types of model \cite{Macy2015}.

In link updating, cooperative agents may sever a link with a defecting neighbor and form a new link. This rule can be interpreted as cooperators punishing defecting neighbors by cutting off social relationships. In this event, one link, $e_{i, j}^{(\lambda)}$, is selected in the same manner as in strategy updating, and a focal agent ($i$ in this example) is also selected. Agent $i$ severs the link with $j$ when $s_i^{(\lambda)} = C$ and $s_j^{(\lambda)} = D$, and then rewires that link to a randomly selected agent; nothing occurs in other combinations of the agents' strategy \cite{Fu2009}. Because accepting a link with a defecting agent does not improve the average payoff, defecting agents cannot rewire their links. 

We conducted Monte Carlo simulations to examine this model. In these simulations, the relaxation process continued $4N \times 10^4$--$ 2N \times 10^6$ periods; subsequently, the sampling process continued $2N \times 10^4$--$5N \times 10^5$ periods. In order to enhance statistical accuracy, we conducted at least ten simulation runs for each combination of parameters. We recorded the mean cooperation frequencies of each layer ($\rho_C^{(1)}$ and $\rho_C^{(2)}$) in order to report simulation results. In addition, we recorded the absolute difference of cooperation frequencies across two layers at each period and calculated the average of these values ($\rho^\Delta _C$). 

\section*{Results and Discussions}
\begin{figure}[tbp]
\centering
\vspace{5mm}
\includegraphics[width = 75mm, trim= 0 15 0 0]{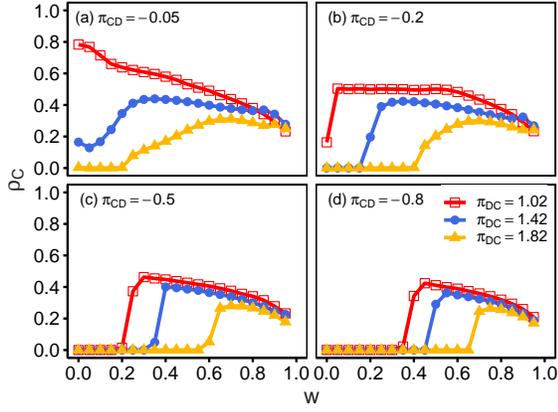}
\caption{\small Frequencies of cooperators ($\rho_C$) are reported as a function of the frequency of link updating ($w$). The introduction of network dynamics (positive $w$) enhances cooperation in many cases but full cooperation ($\rho_C = 1$) is not achieved. Further increases in $w$ decrease cooperation. Other parameters: $N = 1000, p = 0.01,$ and $\mu = 10^{-4}$.}
\label{fig_rho_w}
\end{figure}

The average cooperation frequencies in two layers ($\rho_C = (\rho_C^{(1)} + \rho_C^{(2)})/2$) are given as a function of the frequency of link updating ($w$) to evaluate the overall cooperation level (Figure \ref{fig_rho_w}). With many of the payoff value combinations, the introduction of network dynamics has positive impacts on cooperation. Where $w$ is sufficiently large, the system escapes from almost non-cooperative states. The only exception to this positive impact of link updating is observed when payoff values are advantageous for cooperators ($\pi_{CD} = -0.05$ and $\pi_{DC} = 1.02$).

Although network dynamics increase cooperation up to a point, further increases in $w$ result in decreasing cooperation levels, which has also been observed in one-layer dynamic networks \cite{Fu2009}. Furthermore, cooperation frequency in the network is about 0.5 even with optimal values of $w$. This result contrasts with another model of games run on dynamic duplex networks in which full cooperation was achieved with sufficiently fast link updating \cite{Yang2019d}.

\begin{figure}[tbp]
\centering
\vspace{5mm}
\includegraphics[width = 75mm, trim= 0 15 0 0]{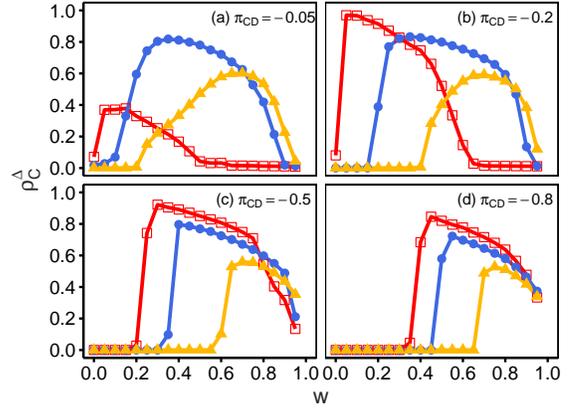}
\caption{\small Absolute differences in the frequencies of cooperators between the two layers ($\rho_C^{\Delta}$) are reported. Moderate values of $w$ induce symmetry breaking as indicated by positive $\rho_C^{\Delta}$. Please refer to Figure \ref{fig_rho_w} for the values of $\pi_{DC}$. Other parameters: $N = 1000, p = 0.01,$ and $\mu = 10^{-4}$.}
\label{fig_dif_w}
\end{figure}

Symmetry breaking across two layers is related to the modest cooperation enhancement shown in Figure \ref{fig_rho_w}. Figure \ref{fig_dif_w} shows the absolute difference in the cooperation frequencies in two layers ($\rho_C^{\Delta}$), which helps to assess if the system shows symmetric behavior. The parameter values in Figure \ref{fig_dif_w} are the same as those in Figure \ref{fig_rho_w}. Two layers show similar cooperation frequencies with fixed edges ($w = 0$) and this pattern holds as long as the values of $w$ remain small. Fixed networks also show asymmetric results with some parameter values. Previous studies have also reported symmetry breaking on static networks \cite{Jin2014, Battiston2017, Liu2019k}. However, as Figure \ref{fig_dif_w} shows, this result occurs only with a limited combination of parameter values and the effects are small compared to those observed with link updating.

With moderate values of $w$, the evolutionary process leads to clear symmetry breaking across two layers. In comparison to Figure \ref{fig_rho_w}, Figure \ref{fig_dif_w} shows that this broken symmetry is accompanied by an increase in $\rho_C$, which demonstrates that cooperation enhancement is uneven across two layers. Because the cooperation frequencies in only one layer increase, the overall enhancement of cooperation is modest. Further increases in $w$, however, result in restored symmetry that corresponds to the diminishing cooperation frequencies shown in Figure \ref{fig_rho_w}.

\begin{figure}[tbp]
\centering
\vspace{5mm}
\includegraphics[width = 75mm, trim= 0 15 0 0]{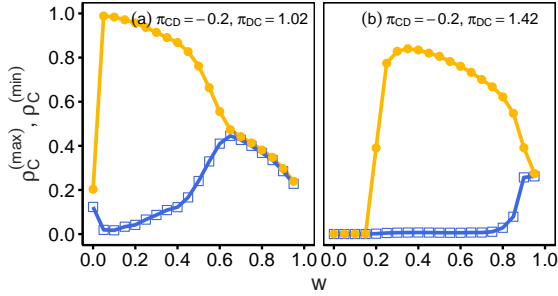}
\caption{\small Cooperation frequencies in two layers are reported separately. $\rho_C^{({\rm max})}$ and $\rho_C^{({\rm min})}$ denote the average frequency of cooperators in a layer that showed higher and lower cooperation frequencies at each period. Link updating ($w > 0$) increases cooperation in one layer while cooperation levels in another layer decrease (panel (a)) or remain the same (panel (b)). Other parameters: $N = 1000, p = 0.01,$ and $\mu = 10^{-4}$.}
\label{fig_rho12_w}
\end{figure}

We also examined the cooperation level in two layers separately to assess the evolutionary outcomes. Here $\rho_C^{({\rm max})}$ and $\rho_C^{({\rm min})}$ denote the average frequency of cooperators in a layer that showed higher and lower cooperation frequencies at each period, respectively. Figure \ref{fig_rho12_w} shows that one layer disproportionately enjoys the benefit of network dynamics. Specifically, panel (a) shows that network dynamics enhance cooperation in one layer at the expense of diminishing cooperation levels in the other layer. This result suggests that network dynamics destabilize the consistent selection of cooperation across two layers. Additionally, panel (b) shows that the introduction of link updating leads to cooperation enhancement in only one layer: cooperation frequencies in the other layer remain almost zero.

\begin{figure}[tbp]
\centering
\vspace{5mm}
\includegraphics[width = 55mm, trim= 0 15 0 0]{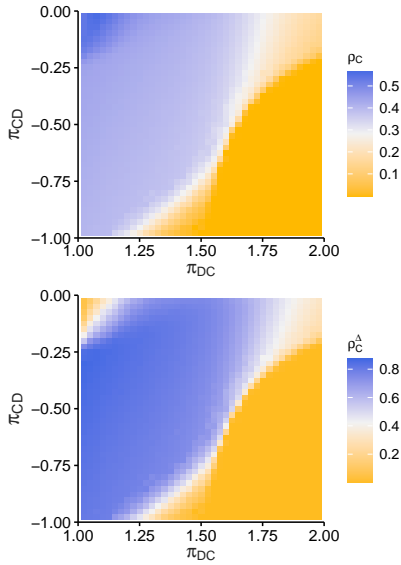}
\caption{\small $\rho_C$ and $\rho_C^{\Delta}$ are reported as a function of two payoff parameters. Lower panel shows broken symmetry (large $\rho_C^{\Delta}$) with a wide range of parameters for which a moderate cooperation level is achieved in upper panel. Other parameters: $N = 1000, p = 0.01, w = 0.5,$ and $\mu = 10^{-4}$.}
\label{fig_phase}
\end{figure}

Broken symmetry is observed with a wide range of payoff values. In Figure \ref{fig_phase}, the value of $w$ was set to 0.5 and the values of two payoff parameters, $\pi_{CD}$ and $\pi_{DC}$, were varied. The upper panel shows that $\rho_C$ is about 0.5 with large $\pi_{CD}$ and small $\pi_{DC}$, whereas small values of $\rho_C$ are observed in the opposite scenario. Between these two scenarios, moderate values of $\rho_C$ are observed, and this region corresponds to large $\rho_C^{\Delta}$, as shown in the lower panel.

\begin{figure}[tbp]
\centering
\vspace{5mm}
\includegraphics[width = 85mm, trim= 0 15 0 0]{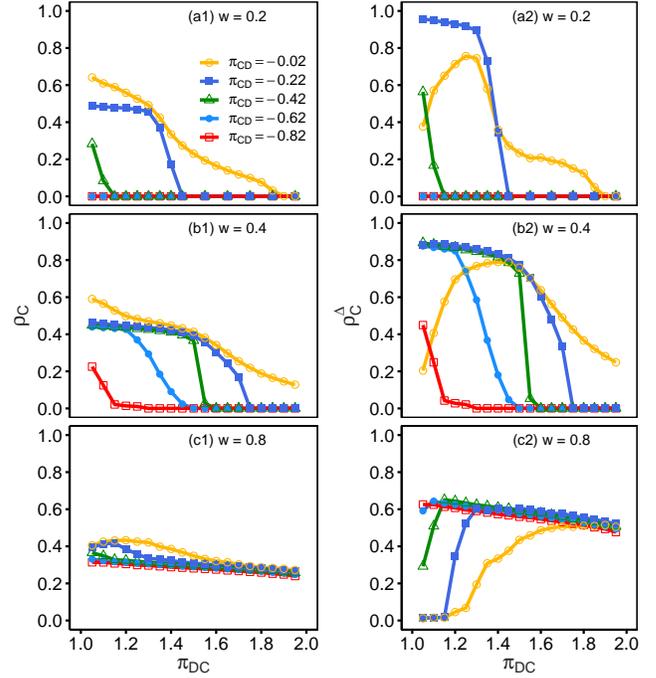}
\caption{\small $\rho_C$ (panels (a1)--(c1)) and $\rho_C^{\Delta}$ (panels (a2)--(c2)) are reported for different values of $w$ and two payoff parameters. $\rho_C^{\Delta}$ shows non-monotonic patterns with large $\pi_{CD}$ (such as $-0.02$). In contrast, $\rho_C$ and $\rho_C^{\Delta}$ decrease monotonically when dilemma strength is large (small $\pi_{CD}$). Other parameters: $N = 1000, p = 0.01, $ and $\mu = 10^{-4}$.}
\label{fig_phase_dif_w}
\end{figure}

We can observe similar patterns with different values of $w$ in Figure \ref{fig_phase_dif_w}. When disadvantage for cooperation is small (i.e., larger $\pi_{CD}$), $\rho_C^{\Delta}$ reaches the peak with moderate values of $\pi_{DC}$. In contrast, as the dilemma harshness increases (i.e., smaller $\pi_{CD}$), $\rho_C^{\Delta}$ decreases monotonically with $\rho_C$. Though payoff values where $\rho_C^{\Delta}$ shows (non-)monotonic patterns depend on the values of $w$, qualitative patterns are similar regardless of the values of $w$.

In addition, we observed positive impacts of link updating. Though exceptions were observed with soft-dilemma situations, fast-link updating supports the survival of cooperation, especially with a harsh social dilemma (small $\pi_{CD}$ and large $\pi_{DC}$). This pattern is consistent with the view that link updating serves as a game-exit option. Notably, this cooperation survival accompanies positive values of $\rho_C^\Delta$, which means that network layers do not show symmetrically enhance cooperation.

\begin{figure}[tbp]
\centering
\vspace{5mm}
\includegraphics[width = 75mm, trim= 0 15 0 0]{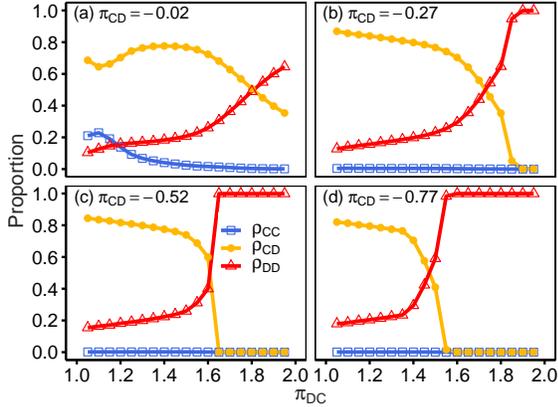}
\caption{\small Strategy selection across two layers is reported. Coherent selection of defection emerges with harsh dilemmas as observed with the large values of $\rho_{DD}$. In contrast, cooperation survival depends on incoherent behavior as indicated by the large (small) values of $\rho_{CD}$ ($\rho_{CC}$). Other parameters: $N = 1000, p = 0.01, w = 0.5$ and $\mu = 10^{-4}$.}
\label{fig_cor _w}
\end{figure}

To further understand this asymmetric state, we examined the strategy correlation between two layers in Figure \ref{fig_cor _w}. The value of $w$ is set to 0.5 as in Figure \ref{fig_phase}. The values of $\rho_{CC}$, $\rho_{CD}$, and $\rho_{DD}$ show the frequencies of agents who adopt cooperation on both, either, or neither layers, respectively. The figure shows that both layers reach non-cooperation with sufficiently harsh dilemma; the value of $\rho_{DD}$ approaches one with sufficiently large $\pi_{DC}$/small $\pi_{CD}$. In contrast, coherent cooperation in both layers is hardly realized. Although $\rho_{CC}$ takes positive values with relatively weak dilemma situations, larger frequencies of $\rho_{CD}$ suggests that cooperation survival can be primarily attributed to the incoherent behavior across two network layers. 

\begin{figure}[tbp]
\centering
\vspace{5mm}
\includegraphics[width = 80mm, trim= 0 15 0 0]{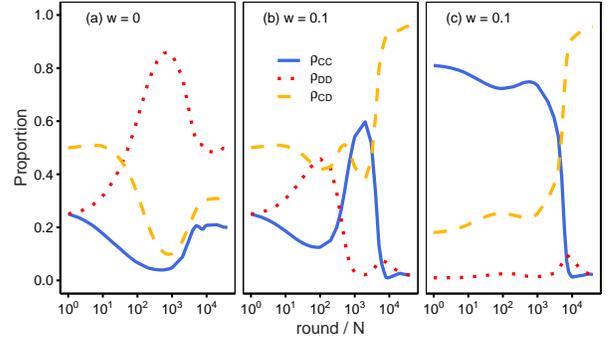}
\caption{\small Time evolution of strategy selection across two layers is reported. Though coherent selection of cooperation (large $\rho_{CC}$) is observed with $w = 0$ (panel (a)), this behavior vanishes once a link-updating event is introduced ($w = 0.1$; panel~(b)). This pattern is observed with different initial cooperation frequencies ($w = 0.1$; panel~(c)). Other parameters: $N = 1000, p = 0.01, \mu = 10^{-4}, \pi_{CD} = -0.2,$ and $\pi_{DC} = 1.02$.}
\label{fig_rho_time}
\end{figure}

To investigate how the system reaches this asymmetrical state, the time evolution of the values of $\rho_{CC}$, $\rho_{CD}$, and $\rho_{DD}$ was assessed and is shown in Figure \ref{fig_rho_time}. Without link updating ($w = 0$), the characteristic pattern observed in evolutionary games on networks is also observed in our model (panel (a)). Cooperation frequency increases due to the formation of cooperative clusters after the initial enduring periods \cite{Wang2013b, Kabir2018}. This pattern is observed with the values of $\rho_{CC}$ and $\rho_{CD}$, which indicate that both layers enjoy the benefit of network reciprocity. 

As shown in Figure \ref{fig_rho_time}, the adoption of cooperation in both layers cannot produce stable outcomes with link updating (panel (b)); the value of $\rho_{CC}$ decreases after reaching its peak. This pattern can be attributed to agents achieving a large \textit{average} payoff if they achieve mutual cooperation in one layer. Although defection in another layer leads to the loss of links with cooperators, interactions in a cooperative layer mainly determine the average payoff as long as degree in that layer is sufficiently larger than degree in a low-cooperation layer. In a previous study that demonstrated full cooperation with fast link updating \cite{Yang2019d}, agents accumulate payoffs from game interactions in two network layers; thus, selecting cooperation and achieving large degree size in both layers contribute to acquiring a large payoff. We surmise that this difference in the payoff-collecting mechanism contributes to the different evolutionary outcomes observed.

The instability of cooperation in both layers can also be investigated using simulation runs with different initial cooperation frequencies. In panel (c) where the initial frequency of cooperation in each layer is 0.9, the frequency of agents who select cooperation on both layers diminishes and asymmetric behavior occupies the population. 

\begin{figure}[tbp]
\centering
\vspace{5mm}
\includegraphics[width = 80mm, trim= 0 15 0 0]{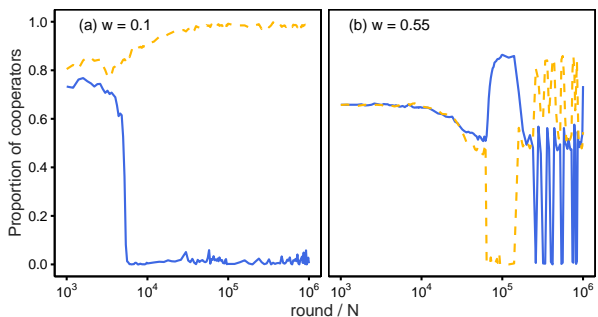}
\caption{\small Time evolution of cooperation frequencies in two layers is reported separately. Each panel shows the result of \textit{one} simulation run. One layer shows stably higher cooperation levels with smaller $w$ (panel (a)) but larger $w$ leads to alternation of cooperative layers (panel (b)). Other parameters: $N = 1000, p = 0.01, \mu = 10^{-4}, \pi_{CD} = -0.2,$ and $\pi_{DC} = 1.02$.}
\label{fig_rho_time_one}
\end{figure}

To further understand the time evolution of the system, Figure \ref{fig_rho_time_one} shows the cooperation level of the two layers in \textit{one} simulation run. Panel (a) reveals that one layer stably shows higher cooperation frequencies with $w = 0.1$, which suggests that positive values of $\rho^\Delta_C$ indicate long-run disparity in cooperation levels between two layers (see the results in panel (b) of Figure \ref{fig_dif_w} which adopts the same parameter values). In contrast, fast link updating ($w = 0.55$) leads to alternation of cooperative layers that betokens the restored symmetry with further large $w$. 

\begin{figure}[tbp]
\centering
\vspace{5mm}
\includegraphics[width = 70mm, trim= 0 15 0 0]{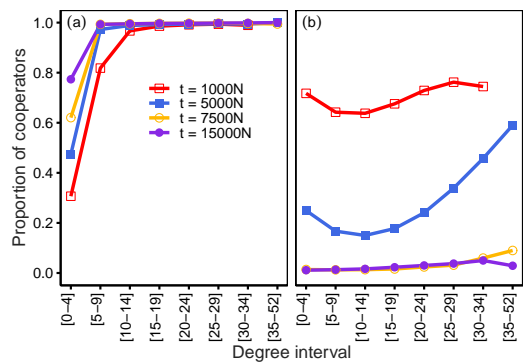}
\caption{\small Frequencies of cooperators are reported as a function of degree in the layer that achieves a \textit{higher} cooperation level at $t = 15000N$. Panel (a) shows that the frequency of cooperators in the high-cooperation layer is positively correlated with degree of that layer. In contrast, panel (b) shows that the frequency of cooperators in the low-cooperation layer and degree in the high-cooperation layer show U-shaped relationships. Other parameters: $N = 1000, p = 0.01, w = 0.1, \mu = 10^{-4}, \pi_{CD} = -0.2,$ and $\pi_{DC} = 1.02$.}
\label{fig_rho_k}
\end{figure}

To investigate how the system reaches this asymmetrical state, we examined the relationship between the nodes' degree and strategy. As shown in Figure \ref{fig_rho_k}, we detected one layer whose cooperation was higher at $t = 15000N$; we then reported the cooperation frequencies of the high-cooperation layer (panel (a)) and a low-cooperation layer (panel (b)) as a function of the degree in the high-cooperation layer. A stylized fact of games on heterogeneous networks, including dynamic networks, is that nodes with a larger degree tend to be cooperative \cite{Santos2006b}. We also confirmed this pattern in our simulations: we observed a positive correlation between cooperation and degree in the high-cooperation layer (panel (a)). Therefore, agents keep choosing cooperation as long as they achieve large degree. 

In contrast, we observed U-shaped relationships between the cooperation frequency in the low-cooperation layer and the degree in the high-cooperation layer (panel (b)). Therefore, cooperators can apparently survive in the low-cooperation layer in two scenarios. First, free-riders who gain a large payoff in the high-cooperation layer can continue to choose cooperation in the low-cooperation layer. This explains the higher cooperation frequency observed among agents whose degree in the high-cooperation layer was small (a small degree suggests that the agents chose defection). Second, agents with a large degree in the high-cooperation layer can remain cooperative because interactions in the low-cooperation layer have a small impact on payoff. Although the cooperation frequency of the groups with the largest degree is noisy due to the small number of observations, these scenarios help to explain the observed relationships. This pattern holds until the cooperation level on the low-cooperation layer approaches zero. 

\begin{figure*}[tbp]
\centering
\vspace{5mm}
\includegraphics[width = 135mm, trim= 0 15 0 0]{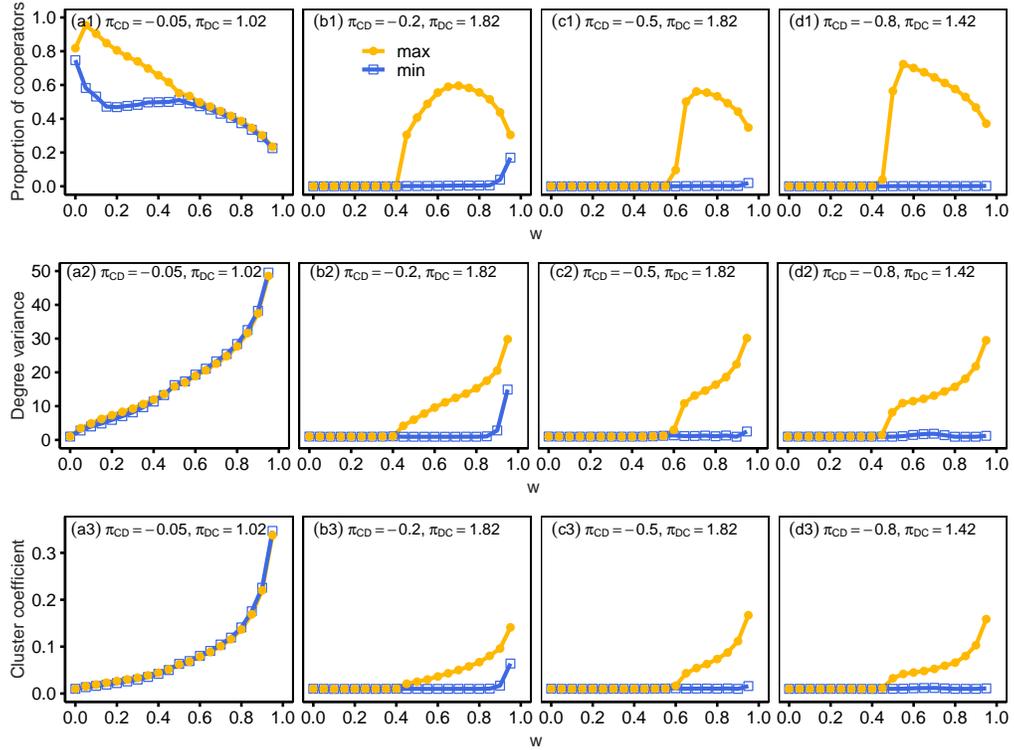}
\caption{\small Cooperation frequencies (panels (a1--d1)), degree variance (panels (a2--d2)), and cluster coefficients (panels (a3--d3)) are reported as a function of the frequency of link updating for different payoff values. Two reported values, max and min, correspond to the values of a layer that achieves large and small cooperation frequency. Discrepancies in cooperation levels do not accompany asymmetrical networks with soft dilemmas and slow link updating (panels (a)). In contrast, harsh dilemmas and fast link updating induce symmetry breaking in the network topology as well as the cooperation levels (panels (b)--(d)). Other parameters: $N = 1000, p = 0.01$ and $\mu = 10^{-4}$.}
\label{fig_topol}
\end{figure*}

Next, we conducted additional simulations and investigated the resultant network characteristics in addition to strategy frequencies. Figure \ref{fig_topol} reposts cooperation frequencies (panels (a1)--(d1)), degree variance normalized by average degree (panels (a2)--(d2)), and cluster coefficients (panels (a3)--(d3)). Two values in the figure, max and min, correspond to the values of a layer that achieves large and small cooperation frequency, respectively. With low dilemma harshness ($\pi_{CD} = -0.05$ and $\pi_{DC} = 1.02$; panels (a)), cooperation frequencies diverge with small $w$. Two layers show similar levels of degree variance and cluster coefficients in these cases, and asymmetry in cooperation does not accompany asymmetrical networks. 

In contrast, a combination of a harsh dilemma and fast link updating induces different cooperation levels and network characteristics. Panels (b)--(d) show that a layer with higher cooperation levels indicates larger degree variance and cluster coefficients. Previous studies have shown that degree heterogeneity and clustering contributes to the evolution of cooperation \cite{Roca2009}. The difference in network topology between two layers agrees with this observation. 

\begin{figure}[tbp]
\centering
\vspace{5mm}
\includegraphics[width = 90mm, trim= 0 15 0 0]{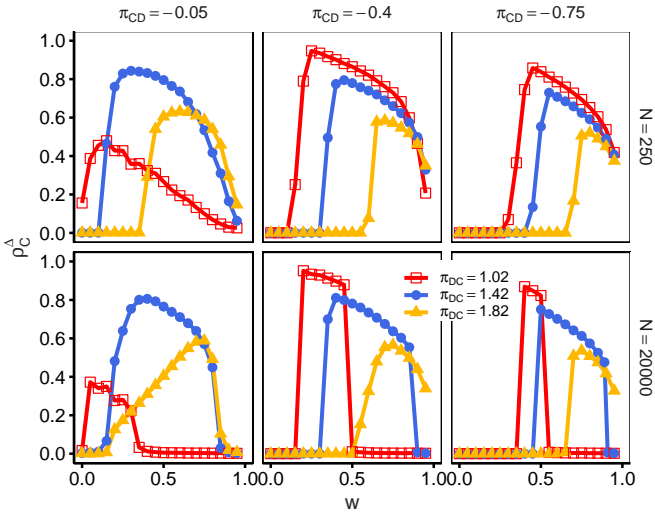}
\caption{\small $\rho_C^{\Delta}$ is reported with different system sizes. Qualitative patterns in Figure \ref{fig_dif_w} are replicated with different system sizes. Value of $p$ is 0.04 and 0.0005 with $N = 250$ and $N = 20000$, respectively. Other parameters: $\mu = 10^{-4}$.}
\label{fig_rho_lm}
\end{figure}

In the rest of this article, we conducted some robustness checks. First, we conducted simulations using a larger ($N = 20000$) or smaller ($N = 250$) network size. Figure \ref{fig_rho_lm} shows $\rho_C^{\Delta}$ as a function of $w$ in the same manner as Figure \ref{fig_dif_w}. As already shown in our network with $N = 1000$, we confirmed that the system shows symmetry breaking with moderate values of $w$ in different sized networks. 

Next, we used different payoff values because the value of $\pi_{CC}$ ($\pi_{DD}$) was set to 1 (0) in the simulation so far. Here, we introduce the framework called universal scaling for the dilemma strength proposed in Ref. \cite{Tanimoto2015c, Wang2015b, Ito2018}. In this framework, harshness of the social dilemma is mainly controlled by two parameters $D_g^\prime$ and $D_r^\prime$, where $D_g^\prime = (\pi_{DC} - \pi_{CC}) / (\pi_{CC} - \pi_{DD})$ and $D_r^\prime = (\pi_{DD} - \pi_{CD}) / (\pi_{CC} - \pi_{DD})$. Larger values of these two parameters indicate harsher dilemma. Payoff values are represented as $\pi_{DC} = \pi_{CC} + (\pi_{CC} - \pi_{DD}) D_g^\prime$ and $\pi_{CD} = \pi_{DD} - (\pi_{CC} - \pi_{DD}) D_r^\prime$ using this framework. The merit of this framework is that the evolutionary outcomes do not depend on the size of $\pi_{CC} - \pi_{DD}$. 

\begin{figure}[tbp]
\centering
\vspace{5mm}
\includegraphics[width = 90mm, trim= 0 15 0 0]{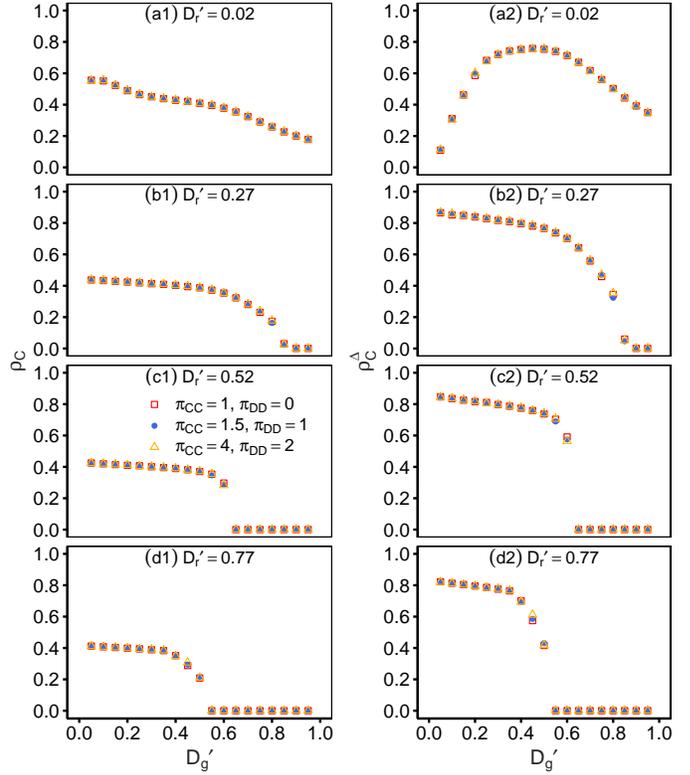}
\caption{\small $\rho_C$ (panels (a1)--(d1)) and $\rho_C^{\Delta}$ (panels (a2)--(d2)) are examined using the framework of universal scaling for the dilemma strength. Outcomes are controlled by $D_g^\prime$ and $D_r^\prime$, and the values of $\pi_{CC}$ and $\pi_{DD}$ have negligible effects. Other parameters: $N = 1000, p = 0.01, w = 0.5$ and $\mu = 10^{-4}$.}
\label{fig_scale}
\end{figure}

Figure \ref{fig_scale} shows $\rho_{C}$ and $\rho_{C}^{\Delta}$ with different values of $\pi_{CC}$ and $\pi_{DD}$. The figure shows that both $\rho_{C}$ and $\rho_{C}^{\Delta}$ do not depend on the values of $\pi_{CC}$ and $\pi_{DD}$. This pattern shows that our findings can be applied with wider range of payoff values. In addition, the result corroborates the validity of the framework of universal scaling in more complex situations, i.e., dynamic multiplex networks.

\begin{figure}[tbp]
\centering
\vspace{5mm}
\includegraphics[width = 75mm, trim= 0 15 0 0]{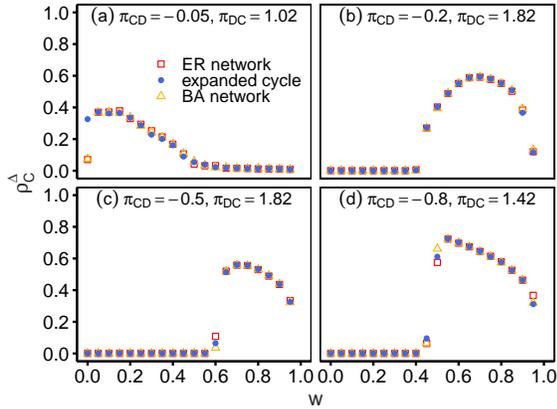}
\caption{\small $\rho_C^{\Delta}$ is examined with different initial networks. Although different network topology affects evolutionary outcomes with fixed networks (e.g., panel (a)), this effect dissipates when network structure coevolves with strategy ($w > 0$). In generating Erd\H{o}s-R\'{e}nyi, expanded cycle, and Barab\'asi-Albert networks, we adopted the parameter values $p = 0.01$, $d = 10$, and $m_0 = m = 5$, respectively. Other parameters: $N = 1000, w = 0.5$ and $\mu = 10^{-4}$.}
\label{fig_init}
\end{figure}

Finally, we examined whether the observed patterns depend on the structure of initial networks. In addition to Erd\H{o}s-R\'{e}nyi networks, we considered expanded cycle, and Barab\'asi-Albert networks \cite{Barabasi1999}. In generating expanded cycles, agents were arranged in a circle, and each agent was connected to $d/2$ neighbors on both sides. In generating Barab\'asi-Albert networks, a complete network with $m_0$ agents was generated; the rest $N - m_0$ agents were connected with $m$ agents following a preferential attachment rule. In the initial states, the two layers were identical: one layer was generated and the other layer was a copy of that generated layer. Figure \ref{fig_init} reports the $\rho_C^{\Delta}$ with different initial network topology. Naturally, networks with different initial topology can result in different outcomes without link updating ($w = 0$, see panel (a)). However, once link updating is introduced ($w > 0$), evolutionary outcomes show similar qualitative patterns regardless of initial networks. This result suggests that the evolution of network topology induced by strategy-based link updating removes the dependency of strategy adoption on initial networks. 

\section*{Conclusion}
Here, we evaluated the evolutionary prisoner's dilemma game played on a dynamic duplex network. The introduction of network dynamics led to enhanced cooperation but the resultant states were far from those of full cooperation. The pattern we observed was related to broken symmetry, whereby link updating led to enhanced cooperation in one layer while cooperation frequencies in another layer remained the same or deteriorated. This state was maintained as long as the frequency of link updating was not overly high. The robust findings of previous studies have demonstrated that network dynamics facilitate cooperation \cite{Perc2010}. Our results show that the ramifications of link updating become more nuanced once network multiplexity is considered. 

Lastly, we consider the potential future extensions of this study. Our simulation results suggest that individuals can show incoherent behavior across multiple social domains: they may choose cooperation in one domain but choose a defecting option in another domain. Future studies could therefore examine under which conditions individuals tend to show (in)coherent behavior in multiple social domains. For example, the present study evaluated average payoff but additional studies might also consider accumulated payoff and a combination of the two methods \cite{Szolnoki2008}. Furthermore, although our study relies on imitation-based evolution (which to date has been widely adopted), other studies have indicated that the strategy-updating rule has a significant role in the evolution of cooperation \cite{Cimini2014}. Studies in these areas may further contribute to our understanding of the conditions under which network multiplexity and network dynamics affect the evolution of cooperation.


\end{document}